\documentclass[12pt]{article}
\usepackage{amsfonts}
\usepackage{amssymb}
\usepackage{graphicx}
\def\siz{\small}
\def\be{\begin{equation}}
\def\ee{\end{equation}}

\begin{document}
\title{NONRELATIVISTIC DARK-ENERGY FLUID IN A BABY UNIVERSE}
\author{P.C. Stichel$^{1)}$  and W.J.
Zakrzewski$^{2)}$
\\
\siz
$^{1)}$An der Krebskuhle 21, D-33619 Bielefeld, Germany \\ \siz
e-mail:peter@physik.uni-bielefeld.de
\\ \\ \siz
$^{2)}$Department of Mathematical Sciences, University of Durham, \\
\siz Durham DH1 3LE, UK \\ \siz
 e-mail: W.J.Zakrzewski@durham.ac.uk
 }

\date{}
\maketitle

\begin{abstract}

We show that the dynamical realization of the acceleration-enlarged Galilean symmetry leads to 
nonrelativistic massless particles whose energy may be negative. We present a fluid mechanical
generalisation of this observation and use it to contruct a nonrelativistic two-dimensional 
fluid model which possesses solutions with a negative energy density. Considering this model
as describing dark energy in a baby universe (two space dimensions) we show that its negative
energy density leads to a repulsive gravitational interaction of the fluid with any test body.

\end{abstract}

\section{Introduction}

From astrophysical observations, it has been known for 10 years that the universe is
undergoing an accelerated expansion \footnote{We will not quote here the large
number of publications which deal with the theoretical and observational
aspects of this subject. Instead we refer only to a very recent review \cite{one} and
the literature quoted therein.}.

According to the Friedmann equations of General Relativity the accelerated
expansion must have its origin either in a positive cosmological constant or in an 
unknown matter component, called dark energy, leading to repulsive gravity
caused by a large negative pressure (cp. \cite{one}).

Another possibility would be modified gravity (cp. \cite{oneb}).
Up to now it has not been 
possible to choose between these different possibilities \cite{two}. A very recent data analysis favours
dark energy over modified gravity [4]. Furthermore, so far,
we have not seen any dark energy model which is motivated by fundamental physics \cite{one}.

In this paper we present a nonrelativistic dark energy model (in two spatial dimensions)
which is based on a symmetry principle: acceleration enlarged Galilean symmetry.
We start by noting that a nonrelativistic (Newtonian) cosmology can be described
by pressureless Friedmann  equations (cp. \cite{three} and the 
literature quoted therein). Thus, within a nonrelativistc framework, repulsive
gravity can be achieved only by introducing some unknown matter with a negative
energy density. For a particle model this is only possible by considering a new kind of massless
particles.
In [6] we introduced a massless nonrelativistic $d=2$ particle model defined by the higher-order Lagrangian
\be
L\,=\,-\frac{\theta}{2}\,\epsilon_{ij}\,\dot x_i \ddot x_j,\quad i,j=1,2.
\label{one}
\ee
In this expression $\theta$ has the dimension of $mass\times time$, {\it ie} (\ref{one}) does not contain any mass-parameter.
Indeed, for, as shown in [7], the Poisson Brackets (PBs) between boost-generators $K_i$ and translation-generators
$P_j$ vanish which implies that (1) describes a nonrelativistic massless particle. 

This existence 
of nonrelativistic massless particles may appear strange at first site; however, as we show in Section 2, the model
based on Lagrangian (1) possesses a modified relation between energy and momentum (or velocity).
Hence, a fluid mechanical generalisation of (\ref{one}) can serve as a model for dark energy in a two-dimensional (`baby') universe, provided that we can find solutions of the model leading to negative energy density. This is indeed the case
as we show in this paper. 

The paper is organised as follows: In Section 2 we review the relevant properties of the particle model 
based on (\ref{one}) and generalise it in the following section to fluid mechanics. In Section 4 we present some
solutions of the corresponding equations of motion (EOM) and discuss their cosmological implications. Finally, in section 5 we mimic the accelerated expansion of the universe by considering the motion of a test particle.
The paper ends with a short section of Conclusions.

\section{Particle Model}
Here we recall from [7] those properties of the model (\ref{one}) which are relevant 
for our discussion.

First of all we note that (\ref{one}) can be rewritten in the 1-st order formalism in which it takes the form
\be
L\,=\,p_i(\dot x_i-y_i)\,-\,\frac{\theta}{2}\,\epsilon_{ij}\,y_i\,\dot y_j
\label{lag}
\ee
and the corresponding EOM are 
\be
\dot x_i\,=\,y_i,\quad \dot p_i\,=\,0,\quad \dot y_i\,=\,\frac{1}{\theta}\,\epsilon_{ij}\,p_j.
\label{eq}
\ee

The Hamiltonian, corresponding to (\ref{lag}) is given by
\be
H\,=\,p_iy_i
\label{H}
\ee
and so can be used to rewrite (\ref{eq}) as Hamiltonian equations 
$\dot Y=\{Y,H\}$ by using the nonvanishing PBs
\be
\{x_i,\,p_j\}\,=\,\delta_{ij},\quad \{y_i,\,y_j\}\,=\,\frac{\epsilon_{ij}}{\theta}
\label{PB}.
\ee

The model possesses various symmetries (acceleration-enlarged Galilean symmetry, conformal symmetry and a further `hidden' $O(2,1)$. All these symmetries were discussed in [7]. Here, we need only conformal symmmetry so we recall that in the
discussion of this symmetry we use in addition to the {\bf time-translation} generator $H$, also
the {\bf dilatation} generator $D$ 
\be 
D\,=\,tH\,-\,x_ip_i
\ee
and the {\bf expansion} generator $K$ 
\be
K\,=\,-t^2H\,+\,2tD\,-\,2\theta\,\epsilon_{ij}\,x_i\,y_j.
\ee
Together, they build the conformal algebra
\be
\{D,\,H\},=\,-H,\quad \{K,\,H\}\,=\,-2D,\quad \{D,\,K\}\,=\,K
\ee
as can be seen by using the PBs (\ref{PB}).

\section{Fluid Dynamics}
\subsection{Lagrange picture}

We generalise $H$ (\ref{H}) and the PBs (\ref{PB}) from the point particle case to the continuum 
case by introducing comoving particle coordinates $\vec \xi\in R^2$, {\it ie} (cp [8]-[10])
\be
x_i(t)\quad \rightarrow\quad x_i(\vec \xi,t),\quad etc
\ee
where, for each value of $\vec \xi$, the variables $x_i(\vec \xi,t)$, $y_i(\vec \xi,t)$ and
$p_i(\vec \xi,t)$ obey the EOM (\ref{eq}).
The Hamiltonian becomes
\be
H\,\quad \rightarrow\quad H\,=\,\int\,d^2\xi\,\,p_i(\vec \xi,t)\,y_i(\vec \xi,t)
\ee
and the PBs become
$$\{x_i(\vec \xi,t),\,p_k(\vec\xi',t)\}\,=\,\delta_{ik}\,\delta^2(\vec\xi-\vec\xi')$$
\be
\{y_i(\vec \xi,t),\,y_k(\vec\xi',t)\}\,=\,\frac{\epsilon_{ik}}{\theta}\,\delta^2(\vec\xi-\vec\xi').
\label{PB2}
\ee
Analogously, the continuum versions of $D$ and $K$ take the form
$$ D\,=\,tH\,-\,\int \,d^2\xi\,\,x_i(\vec \xi,t)\,p_i(\vec \xi,t)$$
\be
K\,=\,-t^2H\,+\,2tD\,-2\theta\epsilon_{ij} \,\int \,d^2\xi\,\,x_i(\vec \xi,t)\,y_j(\vec \xi,t)
\ee
and together with (\ref{PB2}) can be used to determine the infinitesimal conformal transformation 
properties of $x_i(\vec \xi,t)$, $y_i(\vec \xi,t)$ and
$p_i(\vec \xi,t)$.

They are (we suppress the dependence of the fields on $\vec \xi$ and $t$) 
$$ 
\{D,\,x_i\}\,=\,-t\,y_i\,+\,x_i,\quad \{K,\,x_i\}\,=\,-t^2\,y_i\,+\,2tx_i,
$$
\be 
\{D,\,y_i\}\,=\,-\frac{t}{\theta}\epsilon_{ij}\,p_j,\quad \{K,\,y_i\}\,=\,-\frac{t^2}{\theta}\epsilon_{ij}\,p_j\,+\,2x_i,
\ee
$$
\{D,\,p_i\}\,=\,-p_i,\quad \{K,\,p_i\}\,=\,-2tp_i\,-\,2\theta \epsilon_{ij}\,y_j.
$$

\subsection{Eulerian picture}

In the Eulerian picture the dynamics of the fluid is described in terms of $\vec x$ and $t$ 
dependent fields: particle density $\rho(\vec x,t)$, velocity $u_i(\vec x,t)$ and 
momentum $p_i(\vec x,t)$.

Let us note that in standard fluid mechanics we have $p_i(\vec x,t)=m u_i(\vec x,t)$.
In our case these two vector fields will not be parallel to each other (cp [8-10]).

To see this, assuming uniform distribution in $\vec \xi$, we have
\be
\rho(\vec x,t)\,=\, \int\, d^2\xi\,\delta^2(\vec x-\vec x(\vec \xi,t))
\label{ex}
\ee
and
\be
\rho(\vec x,t)\,p_i(\vec x,t)\,=\,\int\,d^2\xi\,p_i(\vec \xi,t)\,\delta^2(\vec x-\vec x(\vec \xi,t))
\label{ex1}
\ee
and an analogous expression for $u_i(\vec x,t)$ (in the expression above replace
$p_i(\vec x,t)$ by $u_i(\vec x,t)$ and $p_i(\vec \xi,t)$ by $y_i(\vec \xi,t)$.
In fact, (\ref{ex1}) holds for any function of relevant variables.

To derive EOM of our fluid dynamics we note that from (\ref{ex}), (\ref{ex1}) and the particle EOM
(\ref{eq}) we have
\begin{itemize}
\item 
\be 
\partial_t\,\rho(\vec x,t)\,+\, \partial_i(\rho u_i)(\vec x,t)\,=\,0, \label{a}
\ee
{\it ie} the continuity equation
\item 
\be 
\partial_t\,u_i(\vec x,t)\,+\,u_k(\vec x,t) \partial_k u_i(\vec x,t)\,=\,\frac{1}{\theta}\,\epsilon_{ij}
p_j(\vec x,t),
\label{b}
\ee
which would be Euler's eq. for a free flow in case of a vanishing r.h.s.
\item
\be 
\partial_t\,p_i(\vec x,t)\,+\,u_k(\vec x,t) \partial_k p_i(\vec x,t)\,=\,0.
\label{c}
\ee
\end{itemize}

With these properties we find that the Hamiltonian $H$ is given by
\be
H\,=\,\int d^2x\,\rho(\vec x,t)\,u_i(\vec x,t)\,p_i(\vec x,t)\ee
and, analogously, $D$ and $K$ are given by
\be
D\,=\,tH\,-\,\int \,d^2x\,\rho(\vec x,t)\,x_i\,p_i(\vec x,t)
\ee
and
\be
K\,=\,-t^2H\,-2tD\,-\,2\theta\,\epsilon_{ij}\,\int\,d^2x\,\rho(\vec x,t)\,x_i\,u_j(\vec x,t).
\ee
These expressions can then be used to determine the infinitesimal conformal transformations of our fields
$\rho(\vec x,t)$, $p_i(\vec x,t)$ and $u_i(\vec x,t)$.

\section{Cosmological Solutions of Fluid Dynamics}
Here we look at applications of our fields in cosmology.
We observe that according to the Newtonian cosmological principle (cp [11], [12]) applied to our $d=2$ model, 
the velocity field $u_i(\vec x,t)$ can only depend linearly on $x_i$ so that
\be
u_i(\vec x,t)\,=\,f_1(t)\,x_i\,+\,f_2(t)\,\epsilon_{ij}\,x_j\ee
leading through the EOM (17) to a corresponding structure for the momentum field $p_i(\vec x,t)$:
\be
p_i(\vec x,t)\,=\,f_3(t)\,x_i\,+\,f_4(t)\,\epsilon_{ij}\,x_j.\label{uu}
\ee
Furthermore, the energy density $E(\vec x,t)$ given by
\be 
E(\vec x,t)\,=\,\rho(\vec x,t)\,u_i(\vec x,t)\,p_i(\vec x,t)
\ee
should be a function of $t$ only; {\it ie}
\be
E(\vec x,t)\,=\,E(t).
\ee

Next we construct solutions of EOM (\ref{a}-\ref{c}) which satisfy these requirements.

\subsection{Conformally Invariant Solutions}

Let us note that $u_i(\vec x,t)$ and $p_i(\vec x,t)$ given by
\be
u_i(\vec x,t)\,=\,\frac{2x_i}{t},\qquad p_i(\vec x,t)\,=\,-\frac{2\theta \,\epsilon_{ij}\,x_i}{t^2}
\ee
are possible solutions of EOM (\ref{a}-\ref{c}). They clearly satisfy the cosmological principle
mentioned above and are conformally invariant. This invariance involves a few steps and, in fact, it is easy to show
that these are the only conformally invariant solutions. However, the energy density, corresponding to these solutions
vanishes and so to have physically interesting solutions we need to break conformal symmetry.

\subsection{Solution breaking scale symmetry}
Let us try to find a solution which break the conformal symmetry in a minimal way; namely, by breaking scale symmetry.
To do this we observe that the particle EOM (\ref{eq}) are invariant with respect to the continuous 
transformations ($\alpha\in R^1$)
$$x_i\quad \rightarrow \quad x_i'\,=\,x_i\,-\,\alpha \,\theta\,\epsilon_{ij}\,y_j$$
\be 
y_i\,\quad \rightarrow \quad y_i'\,=\,y_i\,+\,\alpha \,p_i\label{tr}
\ee
$$ p_i\,\quad \rightarrow \quad p_i'\,=\,p_i.$$
Clearly $\alpha$ has dimension of mass$^{-1}$.

Note that as a result of this transformation the Hamiltonian system (\ref{H}) and (\ref{PB}) goes  over to a different system 
(containing three central charges of the acceleration-enlarged Galilean algebra) as discussed in [13].
To determine the explicit form of the resultant functions $u_i(\vec x,t)$, $p_i(\vec x,t)$ and $\rho(\vec x,t)$
is quite involved as it requires going through the calculations of the effects of the transformations (\ref{tr})
on $u_i(\vec \xi)$, and $p_i(\vec \xi)$. In the end one obtains
\be 
p'_i(\vec x,t)\,=\,\frac{4\alpha tx_i\,-2\,\frac{t^2}{\theta}\,\epsilon_{ij}\,x_j}{4 \alpha^2t^2\,+\,(\frac{t^2}{\theta})^2}
\ee
and
\be
u'_i(\vec x,t)\,=\,\frac{t}{\theta}\,\epsilon_{ij}\,p'_{j}(\vec x,t)\,+\,\alpha\,p'_i(\vec x,t).
\label{aaa}\ee
Note that $p'_i(\vec x,t)$ and $u'_i(\vec x,t)$ both satisfy the cosmological principle mentioned at the beginning 
of this section.

The energy density is now given by
\be
E(\vec x,t)\,=\,\frac{\alpha \rho(\vec x,t)\,r^2}{\alpha^2t^2\,+\,\frac{t^4}{4\theta^2}}.
\ee
In order to get an uniform energy density, as required by the cosmological principle, we seek a solution of the 
continuity eq. (\ref{a}) of the form
\be
\rho(\vec x,t)\,=\,\frac{g(t)}{r^2}.\label{ro}
\ee
However, as $u'_i(\vec x,t)$ is of the form (\ref{uu}) we see that $\dot g(t)=0$ and so $g(t)=g_0$.
With this choice of $\rho(\vec x,t)$, which is scale invariant, we see that the energy density of our
two-dimensional dark-energy fluid is given by
\be 
E(t)\,=\,\frac{g_0\alpha}{\alpha^2t^2\,+\,\frac{t^4}{4\theta^2}},
\ee
and is a negative increasing function of time if $\alpha<0$.

\subsection{A more general solution for $\rho(\vec x,t)$}

The solution (\ref{ro}) for $\rho(\vec x,t)$ leading to a homogeneous energy density, is singular at $r=0$. 
Thus we have to look at more general solutions of the continuity equation (\ref{a}). We require that:
\begin{itemize}
\item 
$\rho(\vec x,t)$ is rotationally invariant, {\it ie}
\be 
\rho(\vec x,t)\,=\,\rho(r,t),
\ee
\item
the velocity field $u_i(\vec x,t)$ is given by (\ref{aaa}) 
{\it ie}
\be 
u_i(\vec x,t)\,=\,f_1(t)\,x_i\,+\,f_2(t)\,\epsilon_{ij}x_i,
\ee
where 
\be 
f_1(t)\,=\,\frac{4\alpha^2t\,+\,\frac{2t^3}{\theta^2}}{4\alpha^2t^2\,+\,(\frac{t^2}{\theta})^2}
\ee
\end{itemize}
Inserting these expressions into (\ref{a}) and noting that we can take $f_2=0$ we see that
\be
\partial_t\rho\,+\,2f_1\rho\,+\,rf_1\partial_r \rho\,=\,0.
\ee
We make an ansatz 
\be
\rho(\vec x,t)\,=\,\frac{1}{r^2}\,g(rF(t))
\ee
and observe that our equation reduces to (for any choice of $g(z)$)
\be
F'(t)\,+\,f_1(t)\,F(t)\,=\,0
\ee
which has a solution
\be
F(t)\,=\,c_1\,\exp(-\int\,dt'\,f_1(t')).\label{int}
\ee

Choosing $g(z)$ of the form 
\be
g(z)\,=\,c_2\frac{z^2}{z^2+a^2}
\ee
we find that $\rho(r,t)$ is given by
\be
\rho(r,t)\,=\,c_2\frac{F^2(t)}{r^2F^2(t)\,+\,a^2}
\ee
leading finally to the energy density
\be
E(r,t)\,=\,\frac{\alpha c_2 F^2(t)\,r^2}{r^2 F^2(t)\,+\,a^2}\,(\alpha^2t^2\,+\,\frac{t^4}{4\theta^2})^{-1}.
\ee

Thus this time $E$ is not homogeneous - but becomes independent of $r$ at large 
scales defined by
\be 
rF>> a.
\ee

The integral (\ref{int}) can also be easily calculated and one finds that
\be
F(t)\,=\,c_1(4\alpha^2t^2\,+\,(\frac{t^2}{\theta})^2)^{-\frac{1}{2}}.
\ee

Thus $F(t)$ and also $E(r,t)$ are singular at $t=0$.

\section{Motion of a test particle}

Here we consider the motion of a rotationally symmetric test particle
of mass $m$ moving in a planar universe filled with baryonic and cold dark matter
of density  $\rho_M(t)$ and a dark energy fluid of energy density
$E(r,t)$. In order to see the essential aspects of the accelerated expansion
in our model we neglect, for the moment, any interaction {\it ie} exchange
of energy, between the dark matter and the dark energy component or the back
reaction of the test particle on the gravitational field.

The particle EOM is then given by
\be
\ddot {\vec{x}}\,=\,-\vec{\nabla}\,\phi(\vec x,t),
\label{EOMa}
\ee
where the gravitational potential is the solution of the Poisson
equation
\be
\nabla^2\,\phi(\vec x,t)\,=\,4\pi G \left(\rho_M(t)\,+\,\frac{E(r,t)}{c^2}\right)
\label{pote}
\ee
with $G$ being the gravitational constant and $c$ the velocity of light.

For the particular case of a homogeneous density $E(t)$ (see (32)) the solution of
(\ref{pote}) is given by
\be
\phi(r,t)\,=\,\pi\,G\,\left(\rho_M(t)\,+\,\frac{E(t)}{c^2}\right)r^2.
\ee

Applying the cosmological principle to the test particle motion we have
[4]
\be
\vec x(t)\,=\, \mu^{-1}\,a(t)\,\vec x(0),
\ee
where $a(t)$ is the cosmic scale factor and $\mu$ is a convenient
rescaling constant. 

The Hubble parameter $H(t)$ is then given by
\be
H=\frac{\dot a}{a}
\ee
and the EOM (\ref{EOMa}) becomes
\be
\frac{\ddot a}{a}\,=\,-2\pi \,G\,\left(\rho_M(t)\,+\,\frac{E(t)}{c^2}\right).
\label{Fr}
\ee
Let us note that (\ref{Fr}) is one of the Friedmann equations in our case.
The other one is the energy conservation (cp [5]).

From (\ref{Fr}) we obtain the accelerated expansion of the universe (as experienced
by our test particle)
if
\be
E(t)\,<\,-c^2\rho_M(t),
\ee
which can always be achieved by an appropriate choice of the constant $\alpha$ in (32).

The generalisation of these considerations to the case of an inhomogeneous $E(r,t)$ 
(see (42)) is straightforward.

\section{Conclusions and Final Remarks}

Dark energy is characterised by possessing repulsive gravity properties. We do not know 
of any physical model which describes it. Hence, maybe, new physical ideas are called
for. Some such attempts have already been made but, so far, to our knowledge, none has been
successful (cp \cite{one}).

In this paper we have shown that the dynamics of dark energy can be based on well
established symmetry principles whose dynamical realisation may be a non-standard one.
To achieve this we have looked at the acceleration-enlarged Galilean 
symmetry which arises from the non-relativistic contraction of the relativistic conformal
symmetry [6]. Particles described by this symmetry necessarily have a vanishing 
rest mass [7] and, as shown in a dynamical model presented in [13], have a non positive-definite
energy. This opens the possibility for the construction of a non-relativistic fluid model
possessing solutions with negative energy density. Such a model could serve then as a dynamical model 
of dark energy. In this paper we have presented such a fluid model in two space 
dimensions. We are currently working on the construction of a similar model in three space dimensions.

{\bf Acknowledgments}: We would like to thank Simon Ross for very helpful comments.

\end{document}